\documentclass[aps,prb,twocolumn,groupedaddress,showpacs]{revtex4}%
\usepackage{graphicx}
\usepackage{amsmath}
\usepackage{amsfonts}
\usepackage{amssymb}%
\providecommand{\U}[1]{\protect\rule{.1in}{.1in}}
\begin{document}
\title{Dislocation networks in helium-4 crystals}
\author{A. D. Fefferman, F. Souris, A. Haziot, J. R. Beamish*, and S. Balibar}
\affiliation{Laboratoire de Physique Statistique de l'ENS, associ\'{e} au CNRS et aux
Universit\'{e}s D. Diderot et P.M. Curie, 24 rue Lhomond 75231 Paris Cedex 05, France}
\affiliation{Permanent address: Department of Physics, University of Alberta, Edmonton,
Alberta, Canada T6G 2E1}
\date{\today }

\begin{abstract}
The mechanical behavior of crystals is dominated by dislocation networks,
their structure and their interactions with impurities or thermal phonons.
However, in classical crystals, networks are usually random with impurities
often forming non-equilibrium clusters when their motion freezes at low
temperature. Helium provides unique advantages for the study of dislocations:
crystals are free of all but isotopic impurities, the concentration of these
can be reduced to the ppb level, and the impurities are mobile at all
temperatures and therefore remain in equilibrium with the dislocations. We
have achieved a comprehensive study of the mechanical response of $^{4}$He
crystals to a driving strain as a function of temperature, frequency and
strain amplitude. The quality of our fits to the complete set of data strongly
supports our assumption of string-like vibrating dislocations. It leads to a
precise determination of the distribution of dislocation network lengths and
to detailed information about the interaction between dislocations and both
thermal phonons and $^{3}$He impurities. The width of the dissipation peak
associated with impurity binding is larger than predicted by a simple Debye
model, and much of this broadening is due to the distribution of network lengths.

\end{abstract}

\maketitle

\section{Introduction}

We have taken advantage of the unique properties of $^{4}$He crystals to
determine the distribution of lengths between nodes in their dislocation
network and to study in detail the interaction between dislocations and both
impurities and thermal phonons. We measured the shear modulus $\mu$ and the
dissipation $Q^{-1}$ associated with dislocation motion since such
measurements provide detailed information about crystal
defects.\cite{Nowick72,Blanter07} The interaction between dislocations and
impurity atoms is responsible for part of the dissipation but its microscopic
mechanism is often controversial in usual materials.\cite{Blanter07,Ritchie82}
Early models\cite{Friedel55,Schoeck63} proposed that the Cottrell atmosphere
of impurities, which immobilizes dislocations at low temperature, can be
dragged at high temperatures. The resulting dissipation produces a relaxation
peak in $Q^{-1}$ that is often broader than expected for a Debye process with
a single relaxation time, but it is difficult to distinguish between the
possible mechanisms for this broadening, e.g., impurity-impurity
interactions\cite{Magalas96} or a distribution of activation energies due to
varying dislocation character.\cite{Numakura91} The $Q^{-1}$ peak width is
also increased by variations in dislocation lengths, but the relevant network
length distribution is very difficult to determine quantitatively in usual
crystals. Transmission electron microscopy (TEM)\cite{Bilde73} and etch pit
images\cite{Lin89} provide some statistical information about dislocation
networks and have been correlated with creep deformation. However, additional
dislocations can be introduced when thin sections are prepared for TEM
studies\cite{Tsui67} and etch pit measurements only probe the dislocation
network at the surface, which may be different from that of the bulk crystal.
Recently, non-destructive measurements of dislocation numbers have been
reported for microcrystals\cite{Jacques13} but these were limited to small
numbers of dislocations.

In hcp $^{4}$He crystals, we have shown that dislocations move over large
distances and produce exceptionally large changes in the elastic shear modulus
$\mu$ and dissipation $Q^{-1}$. Furthermore, the dissipation due to impurity
binding is not obscured by an overlapping dissipation peak due to kink-pair
formation, in contrast to measurements in, e.g., hydrogenated
nickel.\cite{Tanaka83} In $^{4}$He crystals, we detected no effects of the
lattice potential down to 20 mK.\cite{Haziot13a} In addition, the dissipation
peak associated with impurities can be clearly distinguished from the
dissipation from collisions with thermal phonons.\cite{Haziot13b} The only
impurity is $^{3}$He, whose concentration can be lowered far below 1
ppm.\cite{Pantalei10} As a consequence, elastic measurements in $^{4}$He
crystals are extremely sensitive probes of interactions between dislocations
and either highly diluted impurities or thermal phonons. Ultimately, the
behavior of dislocations in $^{4}$He is of fundamental interest because of
quantum effects which could be responsible for the absence of a Peierls
barrier to dislocation motion, allowing them to glide freely. Dislocation
cores may even be superfluid, which would allow \textquotedblleft
superclimb\textquotedblright\cite{Soyler09} in addition to glide.

In this work, we present the distribution $N\left(  L_{N}\right)  $ of lengths
$L_{N}$ between dislocation network nodes in a $^{4}$He crystal and a
comprehensive and quantitative explanation for the effects of phonon and
impurity damping on the vibrations of its dislocations.

\section{Experiment\label{sec:Experiment}}

\begin{figure}[ptb]
\includegraphics[width=\linewidth]{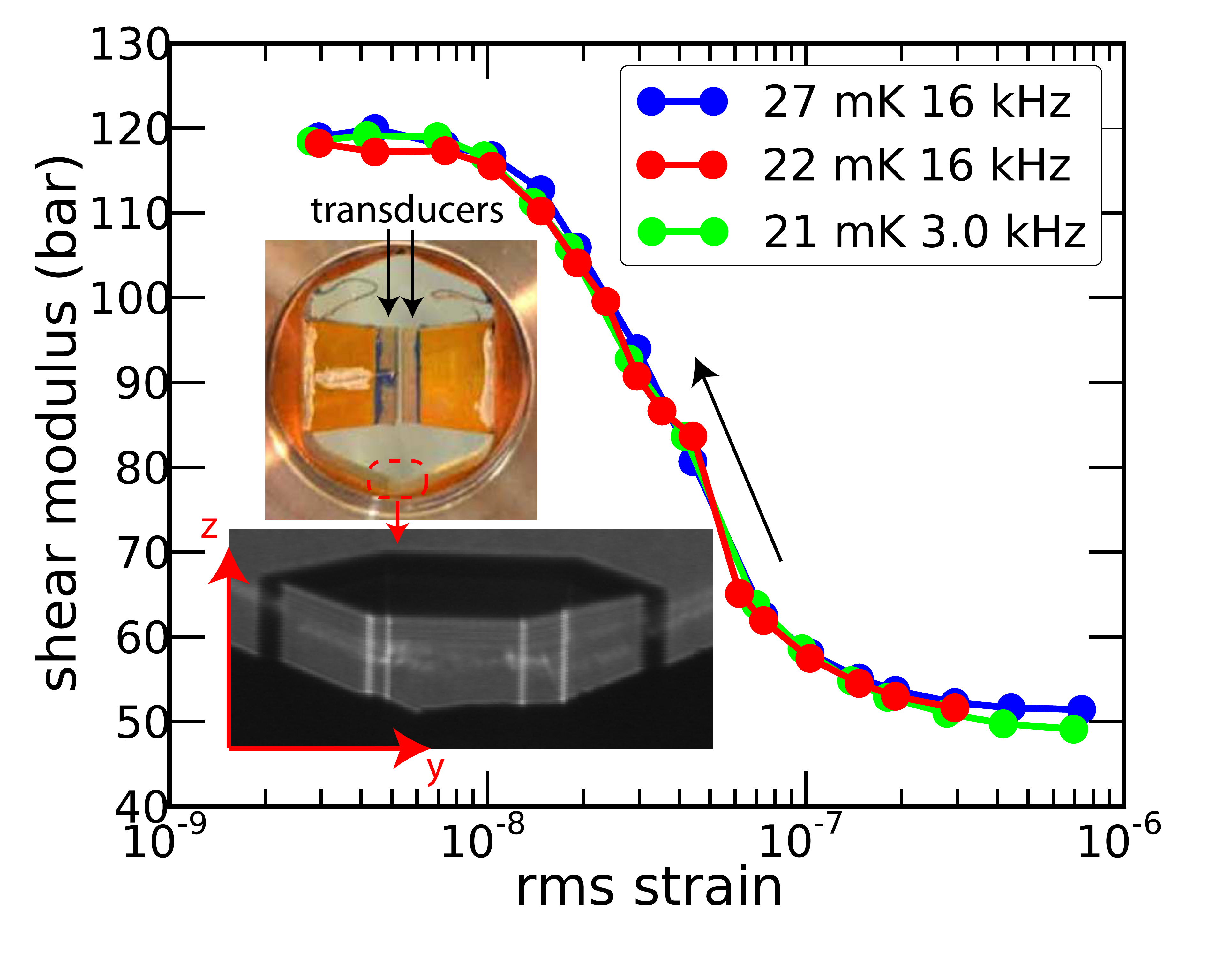}\caption{Three recordings
of the shear modulus measured while decreasing the driving strain near 25 mK.
By analyzing the shape of the transition toward the intrinsic shear modulus
$\mu_{\text{elastic}}$ at low strain, we have determined the width of the
distribution of network lengths. Insets: the experimental cell and the crystal
as it grew from a seed at the bottom of the cell. The transducer plates lie in
planes perpendicular to $y$ and the motion of the left (drive) transducer is
along $z$. }%
\label{fig:mu(epsilon)}%
\end{figure}

We grew single crystals from $^{4}$He with a $^{3}$He concentration
$x_{3}=0.3$~ppm inside a 5~cm$^{3}$ cell, which was made from a hexagonal hole
in a 15 mm thick copper plate closed by two sapphire windows. The orientation
of the crystal was determined from facets that appeared during growth at 600
mK\cite{Balibar05} (Fig. \ref{fig:mu(epsilon)}). In this work, we chose a
crystal with its six fold axis of symmetry close to $z$, with spherical
coordinates $\theta=9.6%
\operatorname{{{}^\circ}}%
$ and $\phi=4.7%
\operatorname{{{}^\circ}}%
$, where $\theta$ is the polar coordinate measured relative to $z$ and $\phi$
is the azimuthal coordinate measured relative to $x$
(Fig.~\ref{fig:mu(epsilon)}). After establishing the orientation of the
crystal, we regrew it from the superfluid liquid at 1.4~K so that any small
liquid regions in the cell were solidified during subsequent cooling. The
temperature of the crystal was then maintained below 1 K so that it remained
completely solid for all of the elastic measurements. This was important
because small liquid regions would have trapped $^{3}$He impurities at low
temperature, leading to temperature variations of the total amount of $^{3}$He
in the crystal.\cite{Pantalei10}

Inside the cell, two piezoelectric shear plates face each other with a
separation $D=0.7$ mm, forming a narrow gap that is filled with the oriented
$^{4}$He crystal (Fig. \ref{fig:mu(epsilon)}). Applying a voltage $V$ to one
transducer produces a shear strain $\epsilon=Vd_{15}/D$ in the $^{4}$He in the
gap, where $d_{15}=0.95$~\AA /V at the temperatures of our
measurements.\cite{Haziot13a} The resulting stress in the $^{4}$He in the gap
is $\sigma=\mu\epsilon$, where $\mu$ is the shear modulus. This stress acts on
the opposite transducer and generates a charge $d_{15}A\sigma$, which is
detected as a current $i\omega d_{15}A\sigma=i\omega d_{15}^{2}\mu VA/D$,
where $A=\left(  1\text{ cm}\right)  ^{2}$ is the area of the transducers and
$\omega$ is the drive frequency. This current was measured using a current
preamplifier connected to a lockin amplifier. In order to determine the
dissipation $Q^{-1}$ in the $^{4}$He, which is the phase difference between
$\sigma$ and $\epsilon$, we subtracted the additional phase shift due to the
measurement electronics from the phase shift between $V$ and the output of the lockin.

The piezoelectric transducers were mounted so that the grounded sides were
facing each other to minimize cross-talk, but there was a residual background
signal due to a capacitive coupling inside the cell, probably between the
wires soldered to the drive and detect transducers. With liquid $^{4}$He in
the cell at a pressure just below the melting curve, this capacitance was
$1.03\times10^{-14}$ F. This is close to the mechanical coupling due to the
shear modulus of the crystal $d_{15}^{2}\mu_{\text{el}}A/D=1.6\times10^{-14}$
F, where $\mu_{\text{el}}=123$ bar is the maximum shear modulus for the
crystal orientation in the present work. We carefully measured the pressure
dependence of the capacitive background with the cell filled with liquid
$^{4}$He and found that it obeys the Clausius-Mosotti equation.\cite{Chan77}
Since the change in the dielectric constant due to the change in density
between the liquid and solid phases at the melting curve is negligible (less
than $1\%$), we subtracted the background measured with liquid just below the
melting pressure $P_{\text{m}}=25.3$ bar from the results of elastic
measurements of the crystal.

\section{Theory}

\subsection{Dislocation Glide}

Dislocations glide in response to stress, allowing the crystal to slip on the
glide plane.\cite{HullChap3} This slip results in a strain $\epsilon_{\text{dis}}$
that adds to the elastic strain $\epsilon_{\text{el}}$ that would be present
in the absence of dislocations, effectively reducing the shear modulus of the
crystal. In hcp $^{4}$He, dislocation glide only reduces the elastic
coefficient $c_{44}$.\cite{Haziot13a} We chose the orientation of the crystal
in the present work so that $\mu$, the component of the elastic tensor that we
measured, was very nearly equal to $c_{44}$ (Appendix A). In the absence of
mobile dislocations and at our working pressure $P_{\text{m}}$, $\mu$ takes
the value $\mu_{\text{el}}=123$ bar.\cite{Greywall77} The effective shear
modulus $\mu=\sigma/(\epsilon_{\text{dis}}+\epsilon_{\text{el}})$ is then
given by%
\begin{equation}
\mu=\frac{\mu_{\text{el}}}{1+\epsilon_{\text{dis}}/\epsilon_{\text{el}}%
}\label{eq:mudef}%
\end{equation}
where $\mu_{\text{el}}=\sigma/\epsilon_{\text{el}}$. The dissipation is
$Q^{-1}=\operatorname{Im}[\mu]/\operatorname{Re}[\mu]$.

In contrast to Zhou \textit{et al}.,\cite{Zhou12,Zhou13} who assumed a large
Peierls barrier against dislocation motion and made predictions that are
inconsistent with experimental data,\cite{Haziot13d} we model the dislocations
as elastic strings\cite{Granato56} that bow out between pinning points in
response to stress over our entire range of temperatures and driving strain.
This model implies the equation of motion\cite{Granato56}%
\begin{equation}
A\ddot{\xi}+B\dot{\xi}-C\frac{\partial^{2}\xi}{\partial x^{2}}=b\sigma
\label{eq:ofmot}%
\end{equation}
where $\xi\left(  x,t\right)  $ is the dislocation displacement as a function
of time $t$ and position $x$ between its pinning points, $A=\pi\rho b^{2}$ is
the dislocation's effective mass per unit length in a material with density
$\rho$, the term in $B$ is the damping force per unit length, the term in $C$
is the effective tension per length in a bowed-out dislocation, and $b\sigma$
is the effective force per length acting on the dislocation. [Because of our
choice of crystal orientation, $\sigma$ is very nearly equal to the resolved
stress $\sigma_{4}$ (Appendix A)]. For an edge dislocation,%
\begin{equation}
C=\left[  \mu_{\text{el}}b^{2}/4\pi\left(  1-\nu\right)  \right]  \left[
\ln\left(  R/r\right)  \right]  \label{eq:defC}%
\end{equation}
where $\nu$ is Poisson's ratio of the material in an isotropic approximation,
$R$ is the distance from the dislocation beyond which its stress field is
canceled by neighboring dislocations (approximately the distance between
dislocations), and $r$ is the dislocation core radius.\cite{Hirth82} In the
present work, $\nu=0.3$, $\rho=191$ kg/m$^{3}$, $b=3.7$ $%
\operatorname{\mathring{A}}%
$, $R\approx100$ $\mu$m, $r\approx1$ nm, $A=8.2\times10^{-17}$ kg/m and
$C=2.2\times10^{-12}$ N.

The resonant frequency of a dislocation of length $L$ between pinning points
is $\omega_{0}=\pi\sqrt{C/A}/L$. It will be shown below that almost all of the
dislocations in our crystal have $L<300$ $\mu$m. Since $\omega_{0}\left(
L=300\text{ }\mu\text{m}\right)  /2\pi\approx300$ kHz, the drive frequency
$\omega$ in the present work is always much less than $\omega_{0}$, so that
the solution of Eq. \ref{eq:ofmot} reduces to%
\begin{equation}
\frac{\xi_{0}}{\sigma}=\frac{16bL^{2}}{\pi^{5}C}\frac{1-i\omega\tau}{1+\left(
\omega\tau\right)  ^{2}} \label{eq:xi_sig}%
\end{equation}
where $\xi_{0}$ is the average oscillation amplitude along the length of the
dislocation and we defined a relaxation time
\begin{equation}
\tau=BL^{2}/\pi^{2}C. \label{eq:tau_def}%
\end{equation}
The slip of the crystal due to glide of a dislocation is $b$ times the
fraction $\xi_{0}L/A_{\text{g}}$ of the glide plane that has slipped, where
$A_{\text{g}}$ is the area of the glide plane. The corresponding contribution
of the dislocation to the strain is $\xi_{0}Lb/V$, where $V$ is the volume of
the crystal. We let $n(L)dL$ be the number of dislocations per unit volume
with pinning length in a differential range around $L$ so that the total
contribution of all the dislocations to the strain is%
\begin{equation}
\epsilon_{\text{dis}}=\frac{b}{2}\int_{0}^{\infty}\xi_{0}\left(  L\right)
Ln(L)dL \label{epsilondistosub}%
\end{equation}
where the factor of 2 comes from averaging over the three possible
$\left\langle 11\bar{2}0\right\rangle $ orientations of the Burgers vector in
the basal plane of our hcp crystal. Substituting Eqs. \ref{eq:xi_sig} and
\ref{eq:defC} in Eq. \ref{epsilondistosub} and using $\mu_{\text{el}}%
=\sigma/\epsilon_{\text{el}}$ yields%
\begin{equation}
\frac{\epsilon_{\text{dis}}}{\epsilon_{\text{el}}}=\alpha\int_{0}^{\infty
}L^{3}\frac{1-i\omega\tau}{1+\left(  \omega\tau\right)  ^{2}}n(L)dL
\label{eq:edis_gen}%
\end{equation}
where $\alpha=32\left(  1-\nu\right)  /\pi^{4}\ln\left(  R/r\right)  $. We
emphasize that $L$ is a generic dislocation pinning length and does not refer
to a particular pinning mechanism. We also note that, in the present work, the
elastic wavelength is greater than 1 cm, so that $\sigma$ and $\epsilon$ are
nearly uniform in the 0.7 mm gap between the transducers.

\subsection{Impurity Pinning\label{sec:impurity_pinning}}

Dislocations in $^{4}$He can be weakly pinned by bound $^{3}$He impurities or
strongly pinned by network nodes. We refer to a dislocation segment between
network nodes as a network link with length $L_{N}$. Impurity binding occurs
because the stress field surrounding a dislocation can lower the elastic
energy associated with an impurity. The concentration of impurities near a
dislocation is modified from the bulk concentration by a factor $\exp\left[
-E_{B}/T\right]  $, where $E_{B}$ is the binding energy.\cite{HullChap10} In solid
$^{4}$He, $^{3}$He impurities are only weakly bound to dislocations, and it is
possible to break a network link away from bound $^{3}$He atoms at high stress.

If one starts at a high oscillating stress amplitude $\sigma$ so that the
network links are free of bound $^{3}$He, the distribution of network lengths
can be inferred from the shear modulus measured while decreasing $\sigma$.
This argument is analogous to the one made by Iwasa in the context of
torsional oscillator experiments.\cite{Iwasa13} As $\sigma$ is decreased,
binding of a $^{3}$He atom to a network link is stable when $\sigma$ reaches a
critical value $\sigma_{c}$. At this point, the bound $^{3}$He atom and the
neighboring network nodes balance the force $\sigma_{c}bL_{N}$ applied to the
network link, and the dislocation does not breakaway from the $^{3}$He atom.
The critical force $F_{c}=\sigma_{c}bL_{N}/2$ is the corresponding force on
the $^{3}$He atom and is determined by the shape of the binding potential. The
bound $^{3}$He atom divides the network link into two shorter segments that,
by the same reasoning, can also be pinned by a $^{3}$He atom at $\sigma_{c}$.
$^{3}$He atoms are available to do so because they are mobile even at zero
temperature.\cite{Allen82} Thus the number of $^{3}$He atoms pinning the
dislocation suddenly increases to the thermal equilibrium value. The
impurity-pinned network link no longer glides very much in response to
$\sigma$, so $\mu$ increases. If there is a distribution of network lengths,
the increase in $\mu$ due to impurity pinning occurs over a range in $\sigma$.

In order to obtain an expression for the distribution of network lengths
$N\left(  L_{N}\right)  $, we consider the case of very low temperatures
$T<<E_{B}$. We first simplify Eq. \ref{eq:edis_gen}. Since $^{3}$He is the
only impurity in our $^{4}$He crystals, the critical force for breaking a
network link away from a single impurity is always the same, and we can define
a critical length $L_{c}=2F_{c}/b\sigma$, which is the length of the network
links that become pinned when the decreasing stress reaches $\sigma$. At
$T<<E_{B}$, the network links with $L_{N}<L_{c}$ are fully pinned by $^{3}$He
and do not contribute to $\epsilon_{\text{dis}}/\epsilon_{\text{el}}$ at all.
Thus we can replace the lower limit of integration in Eq. \ref{eq:edis_gen}
with $L_{c}\left(  \epsilon\right)  $, which we write as a function of
$\epsilon=\sigma/\mu$ because $\epsilon$ is the quantity we directly control.
The network lengths with $L_{N}>L_{c}$ have a length distribution given by
$N\left(  L_{N}\right)  $, which is independent of $\sigma$ because the
network nodes are strong pinning sites. Thus we can also set $L=L_{N}$ and
$n(L)=N\left(  L_{N}\right)  $ in Eq. \ref{eq:edis_gen}, yielding%
\begin{equation}
\frac{\epsilon_{\text{dis}}}{\epsilon_{\text{el}}}=\alpha\int_{L_{c}\left(
\epsilon\right)  }^{\infty}L_{N}^{3}\frac{1-i\omega\tau}{1+\left(  \omega
\tau\right)  ^{2}}N\left(  L_{N}\right)  dL_{N} \label{eq:epdis_interm}%
\end{equation}
The network links with $L_{N}>L_{c}$ are free of bound $^{3}$He atoms, and
measurements at different $^{3}$He concentrations\cite{Haziot13a} demonstrate
that the 300 ppb concentration of unbound $^{3}$He in the present work is too
small to significantly limit $\epsilon_{\text{dis}}/\epsilon_{\text{el}}$.
Damping by thermal phonons is also negligible in the low temperature limit.
Thus we must set $\tau=0$ in Eq. \ref{eq:epdis_interm}:
\begin{equation}
\frac{\epsilon_{\text{dis}}}{\epsilon_{\text{el}}}=\alpha\int_{L_{c}\left(
\epsilon\right)  }^{\infty}L_{N}^{3}N(L_{N})dL_{N} \label{eq:epdis_pin}%
\end{equation}
From Eqs. \ref{eq:epdis_pin} and \ref{eq:mudef} we obtain:
\begin{equation}
N(L_{c})=\frac{1}{\alpha L_{c}^{3}}\left(  \frac{\mu}{\mu_{\text{el}}}\right)
^{-2}\frac{d}{dL_{c}}\left(  \frac{\mu}{\mu_{\text{el}}}\right)
\label{eq:distribution}%
\end{equation}
The distribution of network lengths $N\left(  L_{N}\right)  $ is obtained from
$N(L_{c})$ by renaming the argument.

\section{Results}

\subsection{Strain dependence at low $T$ and high $\omega$}

We measured $\mu$ while decreasing $\epsilon$ to determine the form of
$N\left(  L_{N}\right)  $. In order to prepare for the measurement of $\mu$
shown in Fig. \ref{fig:mu(epsilon)}, we started at 1~K where the crystal is
soft because this is above the $^{3}$He binding energy $E_{B}\approx0.67$ K
(Sec. \ref{sec:3He_damping}). Thus the dislocations were only pinned at
network nodes and vibrated with large amplitudes (Eq. \ref{eq:xi_sig}). We
then cooled the crystal to a temperature near 25~mK under a high strain
$\epsilon=6.8\times10^{-7}$ at $\omega/2\pi>3$ kHz. The crystal remained soft
because the high strain prevented $^{3}$He atoms from binding to the
dislocations, even though the final temperature was well below $E_{B}$.
Holding then the temperature constant, we decreased the applied strain in
steps and measured the equilibrium $\mu$ after waiting 1000 seconds at each
step. We repeated this measurement at the indicated temperatures and
frequencies (Fig.~\ref{fig:mu(epsilon)} legend), starting at 1 K each time,
and $\mu\left(  \epsilon\right)  $ was reproduced to a high degree of accuracy.

The shear modulus gradually increased because $^{3}$He impurities
progressively bound to the dislocations as $\epsilon$ was decreased. If all
the dislocations in our crystal had shared the same network length $L_{N}$
then $\mu$ in Fig.~\ref{fig:mu(epsilon)} would have jumped to $\mu_{\text{el}%
}$ when $\epsilon$ reached a critical strain, as explained in Sec.
\ref{sec:impurity_pinning}. The fact that the transition toward $\mu
_{\text{el}}$ is spread over a rather large range in $\epsilon$ demonstrates
that there is a broad distribution of $L_{N}$. Fig. \ref{fig:mu(epsilon)} also
shows that increasing the temperature from 21 to 27 mK did not cause the
stiffness of the crystal to decrease. The concentration of $^{3}$He bound to
dislocations must have decreased upon increasing the temperature, but the
concentration was still high enough at $T=27$ mK so that the impurity-pinned
dislocations could not move. Thus the measurements of
Fig.~\ref{fig:mu(epsilon)} were made in the low temperature limit assumed in
Section \ref{sec:impurity_pinning}\ and Eq. \ref{eq:distribution} can be used
to determine $N(L_{N})$. To do so we need to convert $\mu\left(
\epsilon\right)  $ (Fig. \ref{fig:mu(epsilon)}) to $\mu\left(  L_{c}\right)
$. This is possible if we know $F_{c}$ since $\epsilon=2F_{c}/bL_{c}\mu$. As
explained below, we determined $F_{c}$ from $\mu\left(  T\right)  $ and
$Q^{-1}\left(  T\right)  $ in the phonon damping regime.

\subsection{Temperature dependence at high $\omega$ and $\epsilon$}

\begin{figure}[ptb]
\includegraphics[width=\linewidth]{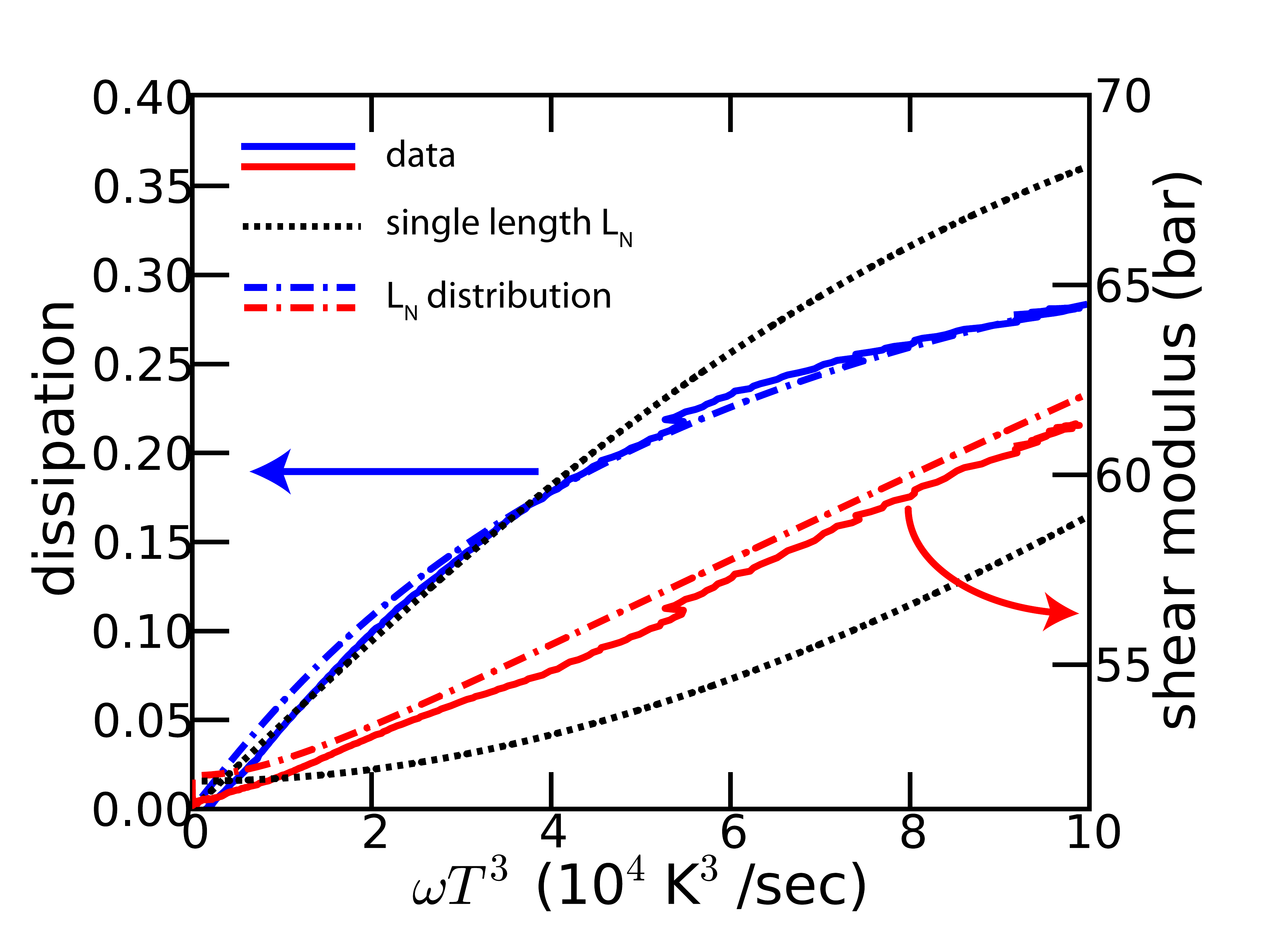}\caption{Solid curves: The
shear modulus $\mu$ and the dissipation $Q^{-1}$ measured while cooling from 1
K with an rms strain $\epsilon= 6.8\times10^{-7}$ at 16~kHz. Dotted curves:
Calculations using a single network length as in our previous
work,\cite{Haziot13b} with best fit values $L_{N_{0}}=96$ $\mu$m and
$\Lambda=7.9\times10^{5}$ cm$^{-2}$. Dash-dotted curves: calculations using
the length distribution from Eq. \ref{eq:distribution} and Fig.
\ref{fig:mu(epsilon)} after adjusting the critical force $F_{c}$ to
$6.8\times10^{-15}$ N.}%
\label{fig:temperature-dependence}%
\end{figure}

Figure \ref{fig:temperature-dependence} shows $\mu$ and $Q^{-1}$ measured
while cooling from 1 K at $\omega/2\pi=16$ kHz. The strain $\epsilon
=6.8\times10^{-7}$ was chosen to be high enough so that no $^{3}$He atoms
could bind to the dislocations as the temperature decreased. Thus the
distribution of pinning lengths $n\left(  L\right)  $ remained equal to the
distribution of network lengths $N\left(  L_{N}\right)  $, and only thermal
phonons could damp the motion of the dislocations. The observed decrease in
$\mu$ and $Q^{-1}$ on cooling occurred because the number of thermal phonons
decreased. After making the substitutions $L=L_{N}$, $n\left(  L\right)
=N\left(  L_{N}\right)  $ and $\tau=\tau_{\text{ph}}$ in Eq. \ref{eq:edis_gen}%
, we obtain:
\begin{equation}
\frac{\epsilon_{\text{dis}}}{\epsilon_{\text{el}}}=\alpha\int_{0}^{\infty
}L_{N}^{3}\frac{1-i\omega\tau_{\text{ph}}}{1+\left(  \omega\tau_{\text{ph}%
}\right)  ^{2}}N(L_{N})dL_{N}\label{edis_ph}%
\end{equation}
with $\tau_{\text{ph}}=B_{\text{ph}}L_{N}^{2}/\pi^{2}C$ from Eq.
\ref{eq:tau_def}. In the low temperature limit, damping occurs because
incident phonons cause the dislocations to \textquotedblleft
flutter\textquotedblright\ and radiate elastic waves,\cite{Ziman01} yielding
$B_{\text{ph}}=14.4k_{B}^{3}T^{3}/\pi^{2}\hbar^{2}c^{3}$ where $c$ is the
Debye sound speed.\cite{Ninomiya74,Ninomiya84} This is the reason why we
plotted $\mu$ and $Q^{-1}$ against $\omega T^{3}\propto\omega\tau_{\text{ph}}$
in Fig. \ref{fig:temperature-dependence}. The $T$ dependence of $\mu$ and
$Q^{-1}$ can be calculated by substituting Eq. \ref{edis_ph} into Eq.
\ref{eq:mudef}. Since $N(L_{N})$ could only be determined from the data in
Fig. \ref{fig:mu(epsilon)} up to the value of $F_{c}$, we take $F_{c}$ as a
fitting parameter. The dash-dot curves in Fig.
\ref{fig:temperature-dependence} show the calculated $\mu\left(  T\right)  $
and $Q^{-1}\left(  T\right)  $ for the best fit value $F_{c}=6.8\times
10^{-15}$~N. We obtain excellent agreement with our measurements over the
entire range in $T$, which supports the form of $N(L_{N})$ determined from the
data in Fig. \ref{fig:mu(epsilon)}.

Haziot et al.\cite{Haziot13b} had fitted $\mu\left(  T\right)  $ and
$Q^{-1}\left(  T\right)  $ in the phonon damping regime assuming a single
length $L_{N_{0}}$. In this case $N(L_{N})=N_{\text{tot}}\delta\left(
L_{N}-L_{N_{0}}\right)  $, where $N_{\text{tot}}$ is the total number of
dislocations per unit volume and $\delta\left(  x\right)  $ is the Dirac delta
function. Substituting this expression for $N(L_{N})$ into Eq. \ref{edis_ph}
yields%
\begin{equation}
\frac{\epsilon_{\text{dis}}}{\epsilon_{\text{el}}}=\alpha\Lambda L_{N_{0}}%
^{2}\frac{1-i\omega\tau_{\text{ph}}}{1+\left(  \omega\tau_{\text{ph}}\right)
^{2}} \label{eq:edis_sing}%
\end{equation}
where $\Lambda=N_{\text{tot}}L_{N_{0}}$ is the total dislocation length per
unit volume. The black dotted curves in Fig. \ref{fig:temperature-dependence}
show $\mu\left(  T\right)  $ and $Q^{-1}\left(  T\right)  $ calculated using
Eq. \ref{eq:edis_sing} with the values $L_{N_{0}}=96$ $\mu$m and
$\Lambda=7.9\times10^{5}$ cm$^{-2}$ chosen to optimize the agreement with the
data in the low temperature limit. These black dotted curves depart from the
data at high temperatures. This is further proof that a broad distribution in
network length is necessary to understand the mechanical properties of $^{4}%
$He crystals.

\begin{figure}[ptb]
\includegraphics[width=\linewidth]{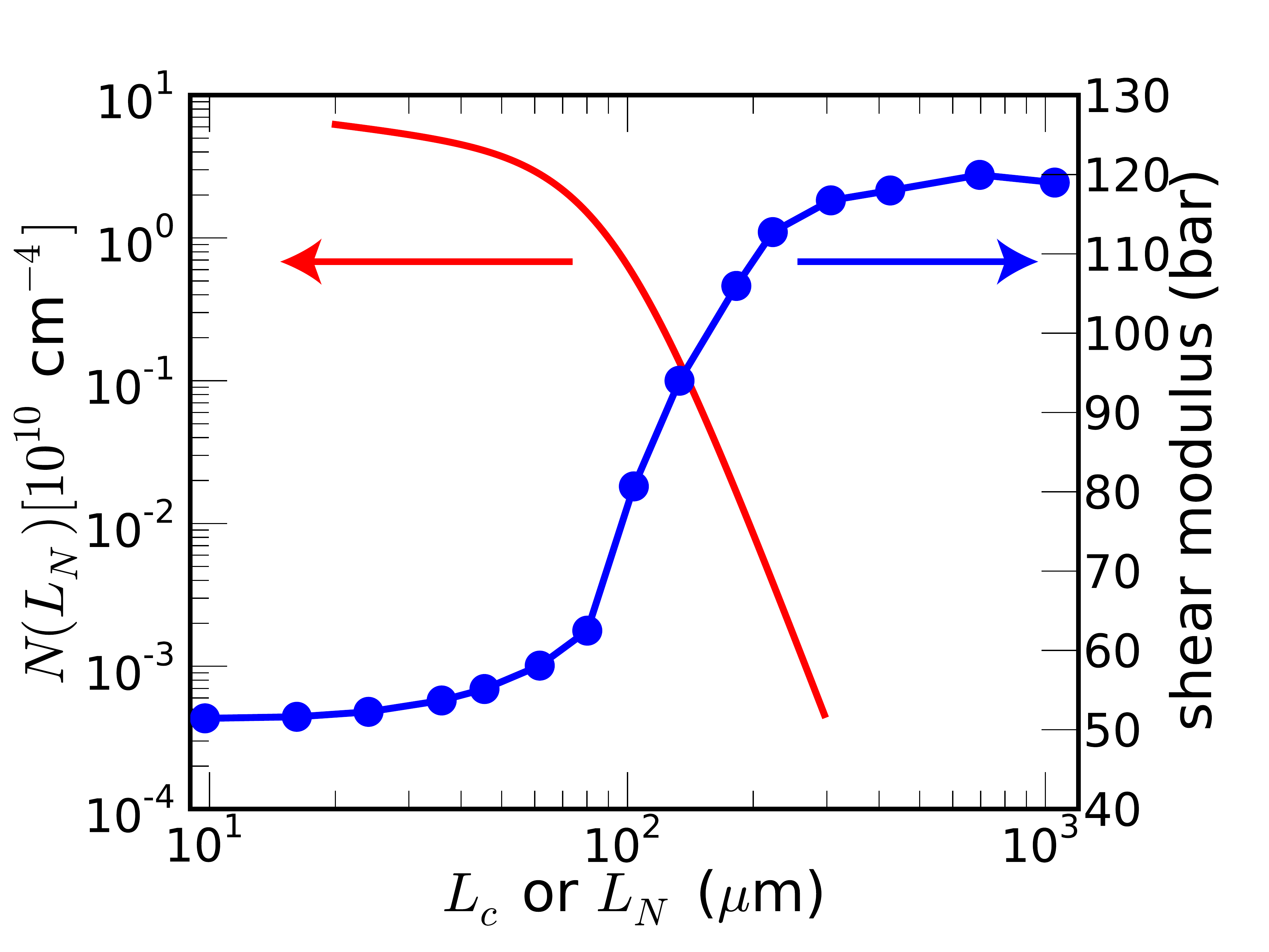}\caption{Blue points: The
shear modulus at 27 mK and 16 kHz from Fig. \ref{fig:mu(epsilon)} as a
function of the critical length $L_{c}=2F_{c}/b\mu\epsilon$ (see text). Red
curve: The distribution $N(L_{N})$ over the range of lengths that contribute
significantly to the softening, derived from the blue points using Eq.
\ref{eq:distribution}}%
\label{fig:distribution}%
\end{figure}

\subsection{Network length distribution}

Having determined $F_{c}$, we have now determined $N(L_{N})$. The blue points
in Fig. \ref{fig:distribution} show the shear modulus data at 27 mK\ and
$\omega/2\pi=16$ kHz from Fig. \ref{fig:mu(epsilon)}, now plotted as a
function of $L_{c}=2F_{c}/b\epsilon\mu$. The red curve is $N(L_{N})$ obtained
from the blue points using Eq. \ref{eq:distribution}. We have only plotted
$N(L_{N})$ over the range $20$ $\mu$m$<L_{N}<300$ $\mu$m corresponding to
dislocations that contribute significantly to the softening, i.e., the range
in $L_{c}$ over which $\mu$ is varying (Fig. \ref{fig:distribution} blue
points). Although there may be a large number of dislocations with $L_{N}<20$
$\mu$m, they do not contribute measurably because of the $L_{N}^{3}$
dependence of $\epsilon_{\text{dis}}/\epsilon_{\text{el}}$ (Eq.
\ref{eq:epdis_pin}). At the opposite extreme, there are very few dislocations
with $L_{N}>300$ $\mu$m. Extending the range of integration beyond $20$ $\mu
$m$<L_{N}<300$ $\mu$m in any of the calculations in the present work does not
affect the results. As expected, the single network length $L_{N_{0}}=96$
$\mu$m obtained from the simplified model of Ref. \onlinecite{Haziot13b} is
near the middle of this range.

\begin{figure}[ptb]
\includegraphics[width=\linewidth]{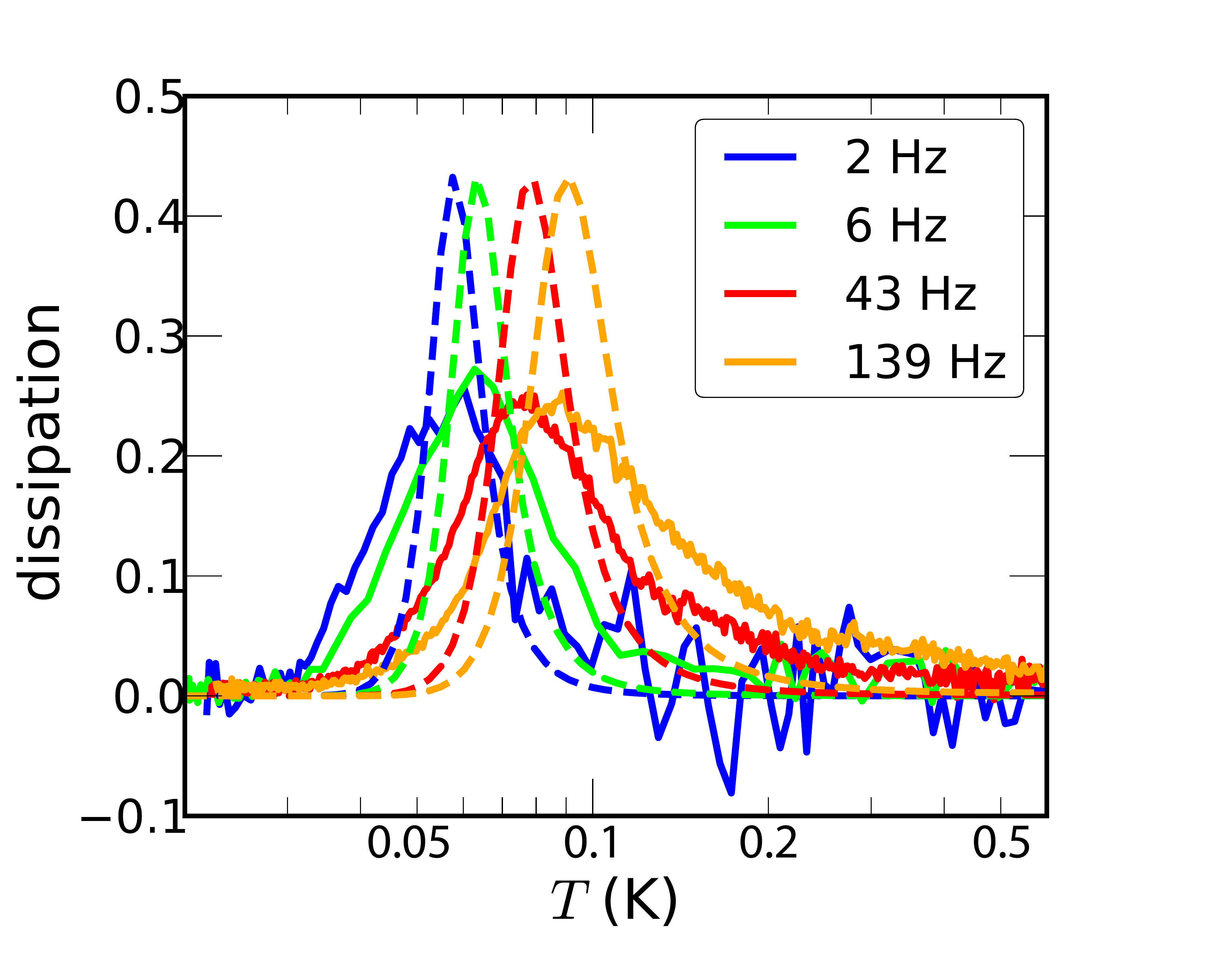}
\includegraphics[width=\linewidth]{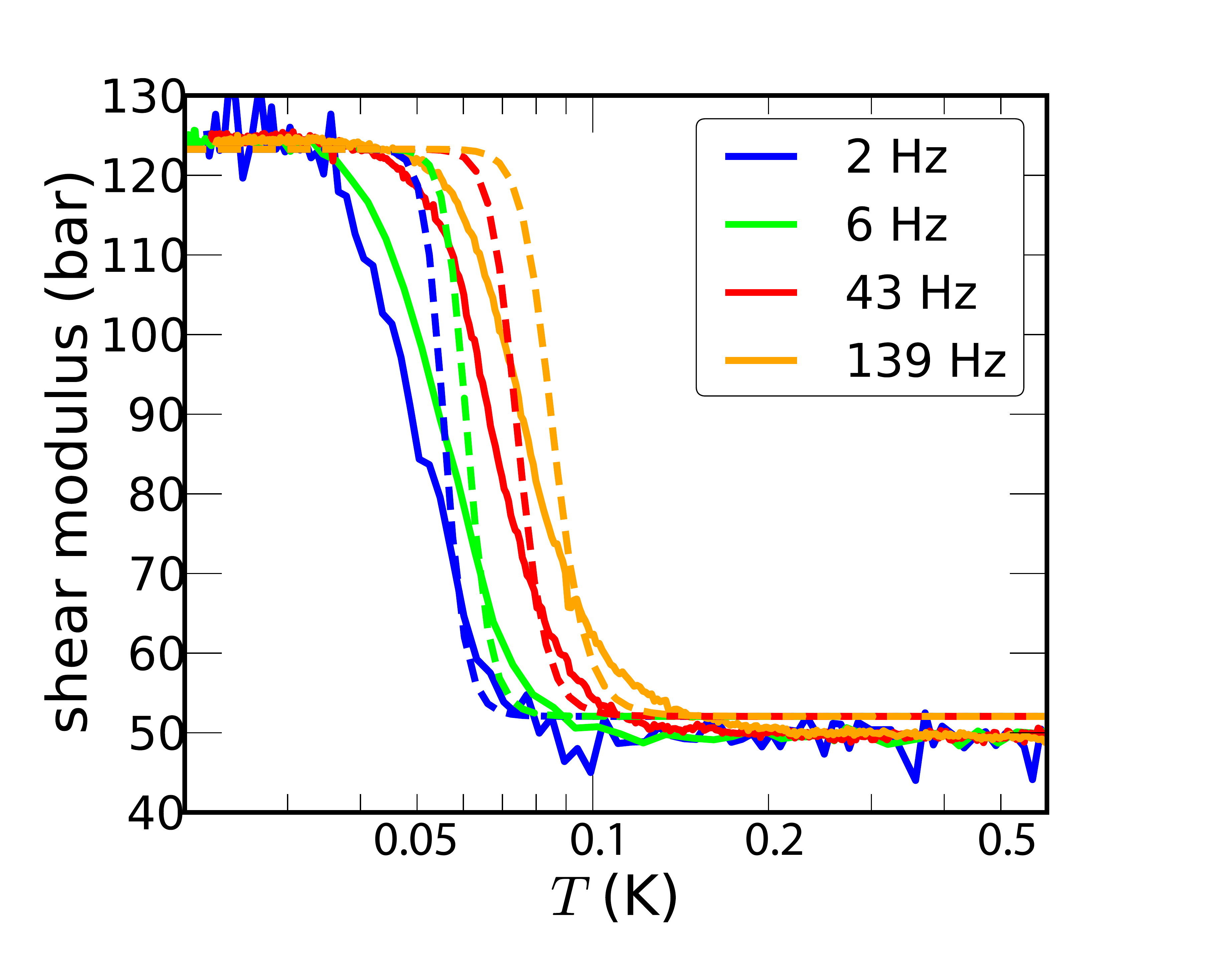}\caption{The dissipation (a)
and the shear modulus (b) measured while cooling from 1 K at low strain
$5.4\times10^{-9}$. Dashed curves: Calculations using a single network length
$L_{N_{0}}=96~\mu$m, a dislocation density $\Lambda=7.9\times10^{5}$ cm$^{-2}%
$, and a single binding energy $E_{B}=0.67$ K.}%
\label{fig:singlen}%
\end{figure}

\begin{figure}[ptb]
\includegraphics[width=\linewidth]{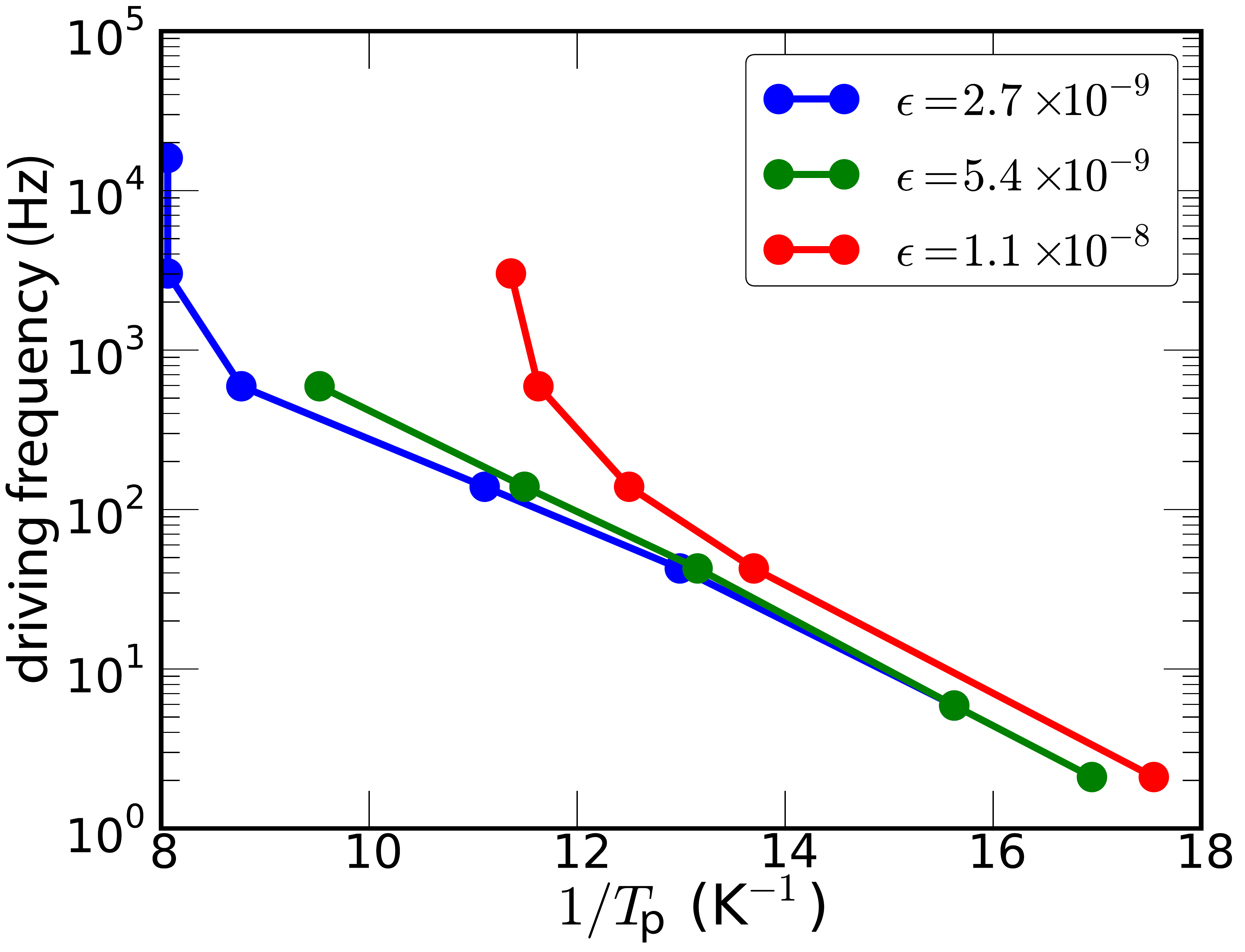}\caption{Arrhenius plot of
the driving frequency versus the inverse temperature of the $Q^{-1}$ peak at
low strain $\epsilon$. This shows the transition from $^{3}$He damping at low
dislocation speeds to $^{3}$He pinning at high dislocation
speeds.\cite{Haziot13c}}%
\label{fig:Arrhenius}%
\end{figure}

\begin{figure}[ptb]
\includegraphics[width=\linewidth]{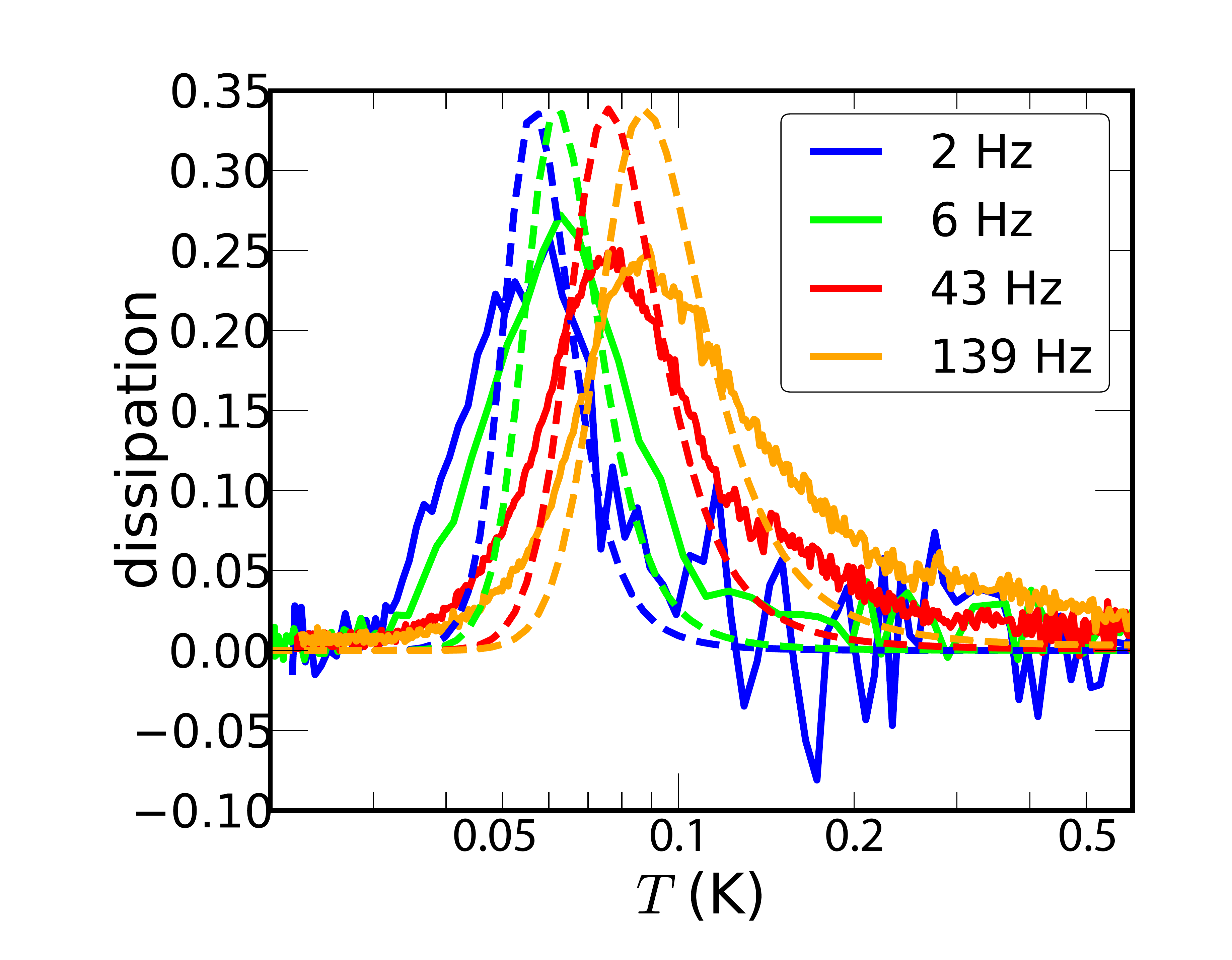}
\includegraphics[width=\linewidth]{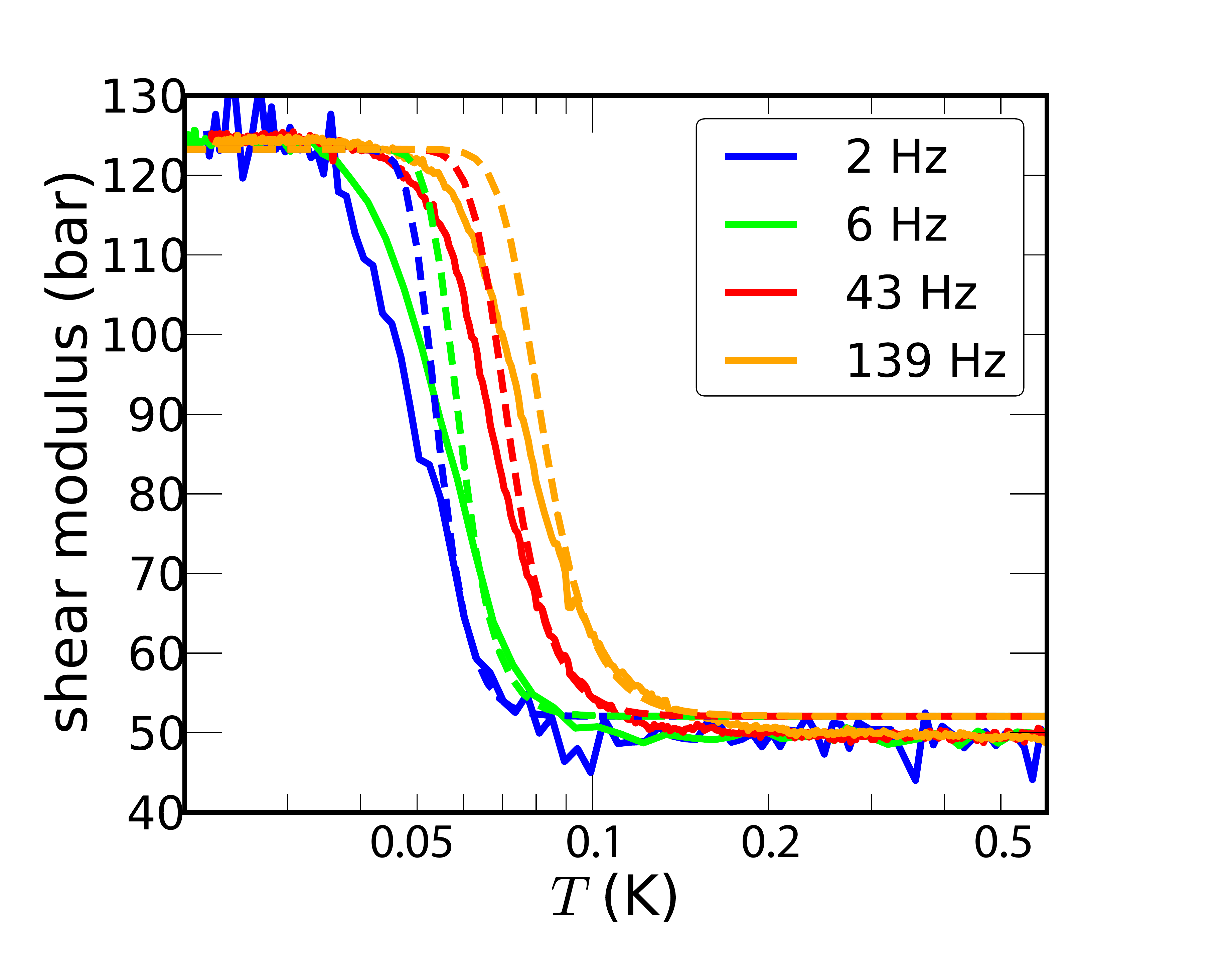}\caption{Same data as in
Fig. \ref{fig:singlen} compared with a calculation (dashed curves) using the
length distribution of Fig. \ref{fig:distribution} and a single
binding energy $E_{B}=0.67$ K.}%
\label{fig:lendist}%
\end{figure}

\begin{figure}[ptb]
\includegraphics[width=\linewidth]{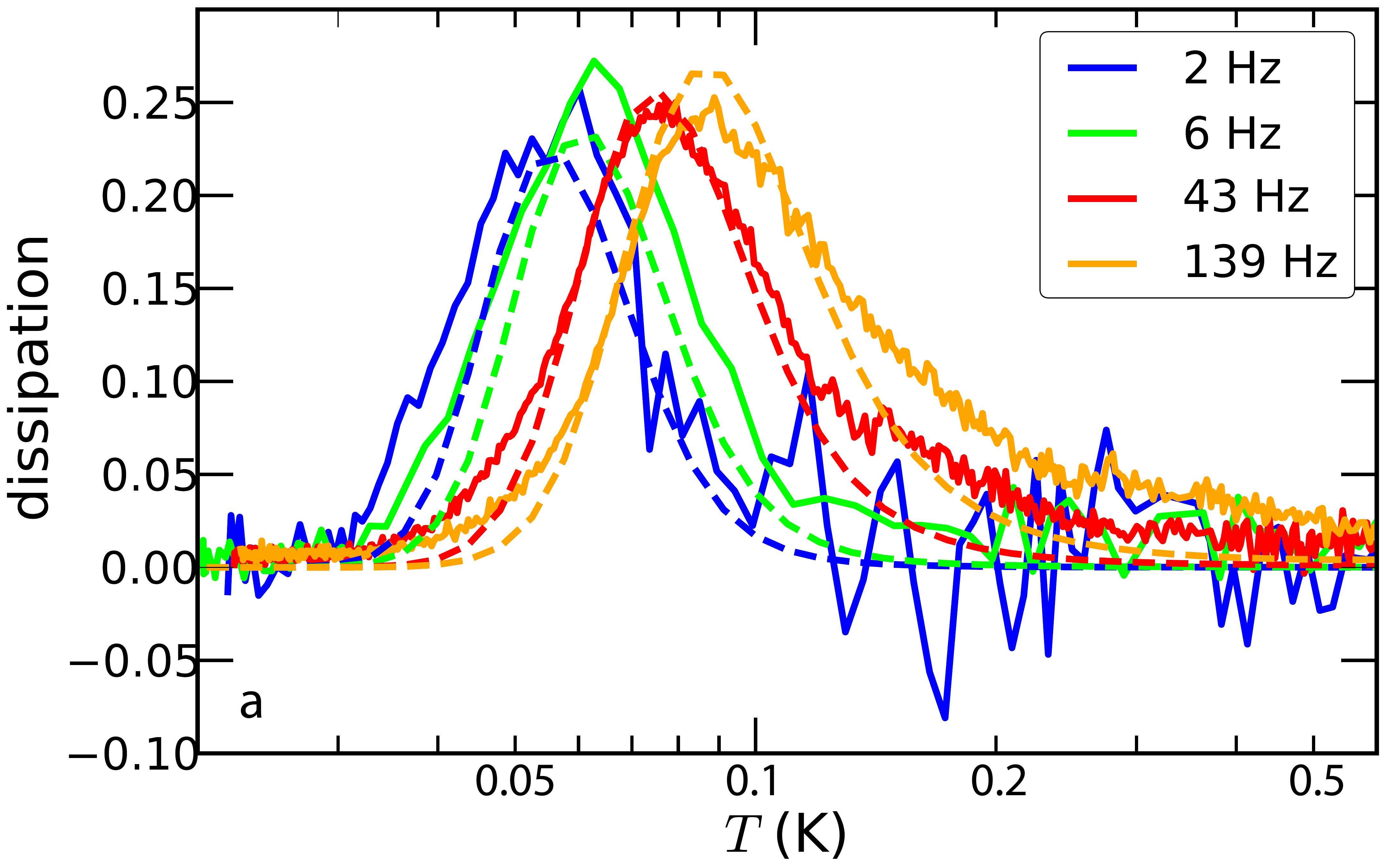}
\includegraphics[width=\linewidth]{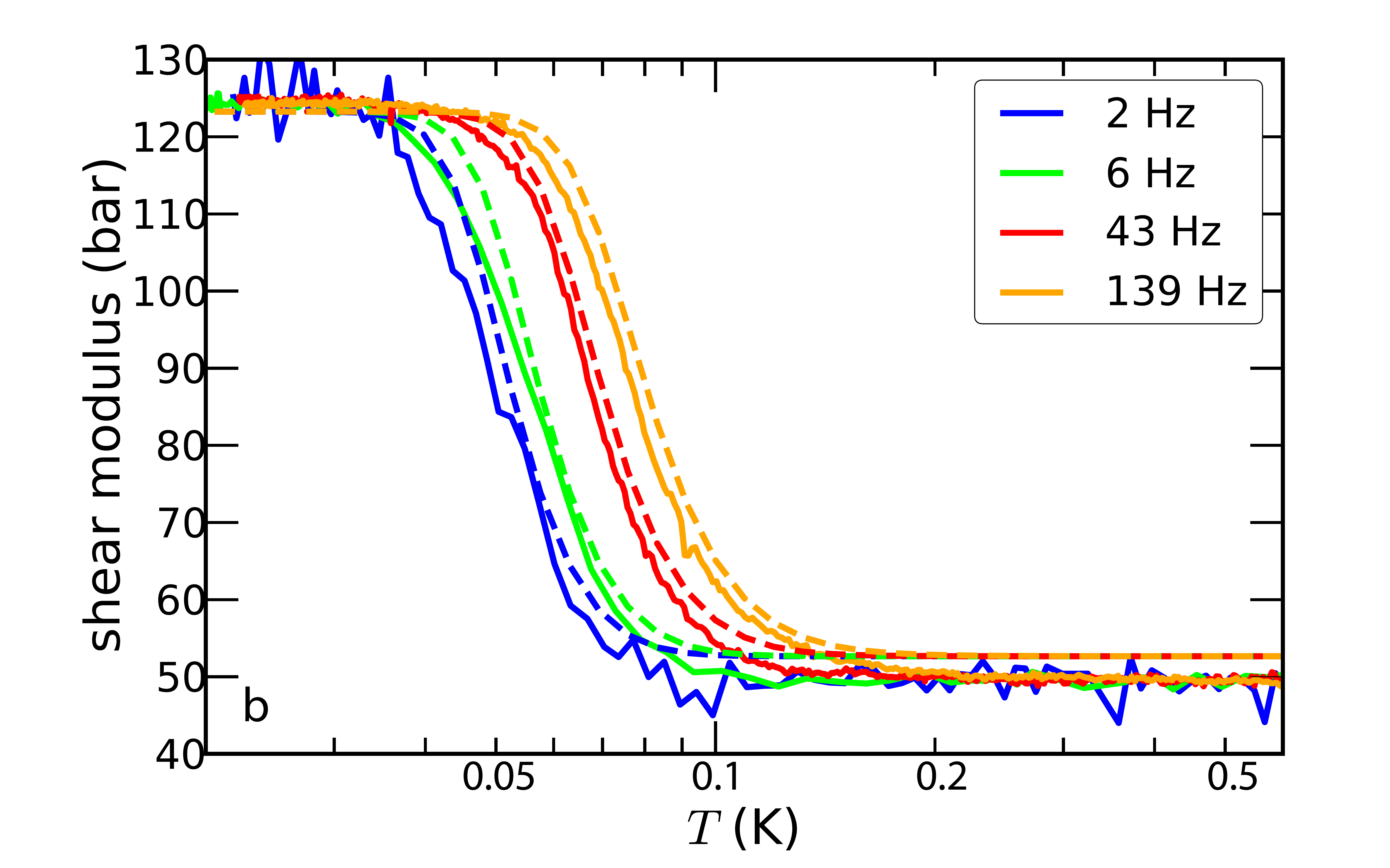}\caption{Same data as in
Fig. \ref{fig:singlen} compared with a calculation (dashed curves) using the
length distribution of Fig. \ref{fig:distribution} and a log-normal
distribution of binding energies $E_{B}$ with mean 0.67 K and standard
deviation 0.1 K.}%
\label{fig:helium3}%
\end{figure}

\subsection{Temperature dependence at low $\omega$ and $\epsilon
$\label{sec:3He_damping}}

Given these results, it is now possible to analyze the effect of damping by
bound $^{3}$He impurities. The data in Fig. \ref{fig:mu(epsilon)} were
acquired in the low temperature limit, so that dislocations with bound $^{3}%
$He atoms did not contribute to $\epsilon_{\text{dis}}/\epsilon_{\text{el}}$.
Now we present a temperature dependent measurement and study the transition
due to thermal binding of $^{3}$He to the dislocations. Figure
\ref{fig:singlen} shows $\mu\left(  T\right)  $ and $Q^{-1}\left(  T\right)  $
measured on cooling at low frequencies and low strain $\epsilon=5.4\times
10^{-9}$. At high temperatures, at least up to 1 K, the crystal is soft with a temperature
independent $\mu$, indicating that damping of dislocation motion by thermal
phonons is negligible at these drive frequencies. The high temperature value
of $\mu$ in Fig. \ref{fig:singlen} is the same as the low temperature value of
$\mu$ in Fig. \ref{fig:temperature-dependence} since in both cases the
dislocations vibrate freely between network nodes. The difference in the temperatures 
at which the dislocation motion is unaffected by both thermal phonons and $^3$He comes from the difference 
in the drive amplitude and frequencies. At low temperatures in
Fig. \ref{fig:singlen}, $\mu$ reaches the intrinsic value $\mu_{\text{el}}$.
The behavior of $\mu\left(  T\right)  $ and $Q^{-1}\left(  T\right)  $ at
intermediate temperatures is frequency dependent: At each frequency
$\omega/2\pi$, we observed a $Q^{-1}$ peak at a temperature $T_{\text{p}%
}\left(  \omega\right)  $ nearly coincident with the midpoint of the
transition in $\mu$, and $T_{\text{p}}\left(  \omega\right)  $ increases with
$\omega$.

We expect $\epsilon_{\text{dis}}/\epsilon_{\text{el}}$ in these measurements
to be given by Eq. \ref{eq:edis_gen}. In this equation, binding of $^{3}$He to
dislocations can change $\epsilon_{\text{dis}}/\epsilon_{\text{el}}$ by
changing $n\left(  L\right)  $ or $\tau$. The frequency dependence of
$T_{\text{p}}$ suggests that changes in $\tau$ dominate in the measurements of
Fig. \ref{fig:singlen}. Figure \ref{fig:Arrhenius} shows $\log\omega$ plotted
against $1/T_{\text{p}}\left(  \omega\right)  $ for the measurements shown in
Fig. \ref{fig:singlen} as well as additional measurements of this type on the
same crystal but at different $\epsilon$ and $\omega$. The measurements at
$\epsilon=2.7\times10^{-9}$ show a transition between frequency dependent
$T_{\text{p}}$ at small $\omega$ and frequency independent $T_{\text{p}}$ at
large $\omega$. The measurements at $\epsilon=1.1\times10^{-8}$ show the same
transition, but at a smaller $\omega$. The transition in fact occurs at a
critical dislocation speed.\cite{Haziot13c} Below the critical speed, changes
in $\epsilon_{\text{dis}}/\epsilon_{\text{el}}$ result from changes in $\tau$,
while above the critical speed, changes in $\epsilon_{\text{dis}}%
/\epsilon_{\text{el}}$ result from changes in $n\left(  L\right)  $. This
implies that below the critical speed $^{3}$He atoms move with the
dislocations and damp their motion, while above the critical speed they
approximate static pinning sites. Figure \ref{fig:Arrhenius} shows that the
measurements in Fig. \ref{fig:singlen} were made well below the critical
dislocation speed. Thus for these measurements, the distribution of pinning
lengths remains equal to the distribution of network lengths as the crystal is
cooled from 1 K. Making the substitutions $L=L_{N}$, $n\left(  L\right)
=N\left(  L_{N}\right)  $ and $\tau=\tau_{\text{3}}$ in Eq. \ref{eq:edis_gen}
yields%
\begin{equation}
\frac{\epsilon_{\text{dis}}}{\epsilon_{\text{el}}}=\alpha\int_{0}^{\infty
}L_{N}^{3}\frac{1-i\omega\tau_{\text{3}}}{1+\left(  \omega\tau_{3}\right)
^{2}}N\left(  L_{N}\right)  dL_{N}\label{eq:3He_damp}%
\end{equation}
where $\tau_{3}=B_{3}L_{N}^{2}/\pi^{2}C$ from Eq. \ref{eq:tau_def}. We assume,
as in previous work,\cite{Iwasa10} that the $^{3}$He damping force is
proportional to the concentration of $^{3}$He bound to the dislocations, so
that the damping constant has the form $B_{3}=B_{3}^{\infty}\exp[E_{B}/T]$.

We can estimate the binding energy $E_{B}$ by considering the case of a single
network length $L_{N_{0}}$. As in the derivation of Eq. \ref{eq:edis_sing}, we
substitute $N(L_{N})=N_{\text{tot}}\delta\left(  L_{N}-L_{N_{0}}\right)  $
into Eq. \ref{eq:3He_damp} to obtain%
\begin{equation}
\frac{\epsilon_{\text{dis}}}{\epsilon_{\text{el}}}=\alpha\Lambda L_{N_{0}}%
^{2}\frac{1-i\omega\tau_{3}}{1+\left(  \omega\tau_{3}\right)  ^{2}}
\label{eq:edis_sing_3He}%
\end{equation}
It is shown in Appendix B that Eq. \ref{eq:edis_sing_3He} implies the
following Arrhenius equation relating $T_{p}$ and $\omega$:%
\begin{equation}
\ln\omega=\ln\left(  \frac{\sqrt{1+s}}{\tau_{0}}\right)  -E_{B}/T_{p}
\label{eq:Arrhen_law}%
\end{equation}
where $s=\alpha\Lambda L_{N_{0}}^{2}$ and $\tau_{0}=B_{3}^{\infty}L_{N}%
^{2}/\pi^{2}C$. We fitted Eq. \ref{eq:Arrhen_law} to the low frequency, linear
part of the data obtained at $\epsilon=2.7\times10^{-9}$ in Fig.
\ref{fig:Arrhenius} in order to obtain the initial estimates $E_{B}=0.67$ K
and $B_{3}^{\infty}=2.6\times10^{-9}$ N sec/m$^{2}$ for use in the subsequent calculations.

In Fig. \ref{fig:singlen}, we show $\mu\left(  T\right)  $ and $Q^{-1}\left(
T\right)  $ calculated using Eqs. \ref{eq:edis_sing_3He} and \ref{eq:mudef}
with a single network length $L_{N_{0}}=96$ $\mu$m and dislocation density
$\Lambda=7.9\times10^{5}$ cm$^{-2}$ (from Fig.
\ref{fig:temperature-dependence}), a single binding energy $E_{B}=0.67$ K, and
a damping coefficient $B_{3}^{\infty}=2.6\times10^{-9}$ N sec/m$^{2}$. In Fig.
\ref{fig:lendist}, we show $\mu\left(  T\right)  $ and $Q^{-1}\left(
T\right)  $ calculated using Eqs. \ref{eq:3He_damp} and \ref{eq:mudef} with
$N\left(  L_{N}\right)  $ from Fig. \ref{fig:distribution} and the same values
of $E_{B}$ and $B_{3}^{\infty}$. The calculated and measured $Q^{-1}$ peak
temperatures $T_{\text{p}}\left(  \omega\right)  $ are in good agreement in
both Fig. \ref{fig:singlen} and Fig. \ref{fig:lendist}. The overall agreement
between theory and data is better in Fig. \ref{fig:lendist} than in Fig.
\ref{fig:singlen}, but it is not perfect in Fig. \ref{fig:lendist}: the
magnitude and temperature width of the $Q^{-1}$ peak were respectively 140\%
and 64\% of the measured values. In order to obtain an excellent fit, we had
to account for the distribution in binding energies that was first proposed in
Ref. \onlinecite{Syshchenko10}.

One expects such a distribution because $E_{B}$ depends on the screw/edge
character of the dislocations,\cite{HullChap4} and variations in the dislocation
character were observed in X-ray images of subboundaries of $^{4}$He
crystals.\cite{Iwasa95} The large softening that we observed, which implies
large $\Lambda L_{N}^{2}$, requires the existence of such
subboundaries.\cite{Haziot13b} Thus we add a distribution of binding energies
$p\left(  E_{B}\right)  $ to Eq. \ref{eq:3He_damp}:
\[
\frac{\epsilon_{\text{dis}}}{\epsilon_{\text{el}}}=\alpha\int_{0}^{\infty
}L_{N}^{3}\int_{0}^{\infty}\frac{1-i\omega\tau_{3}}{1+\left(  \omega\tau
_{3}\right)  ^{2}}p\left(  E_{B}\right)  N(L_{N})dE_{B}dL_{N}%
\]
The distribution $p\left(  E_{B}\right)  $ must have an upper cutoff. In order
to effectively satisfy this constraint and to facilitate comparison with
earlier work,\cite{Syshchenko10} we choose a log-normal form:%
\begin{equation}
p\left(  E_{B}\right)  =\frac{\exp\left[  -\left(  \ln E_{B}-m\right)
^{2}/s^{2}\right]  }{\sqrt{\pi}sE_{B}}%
\end{equation}
with mean $\left\langle E_{B}\right\rangle =\exp\left[  m+s^{2}/4\right]  $
and variance $\left\langle E_{B}\right\rangle ^{2}\left[  \exp\left(
s^{2}/2\right)  -1\right]  $. We obtain an excellent fit to the data with
$\left\langle E_{B}\right\rangle =0.67$ K and a standard deviation of $0.1$ K
(dashed curves Fig. \ref{fig:helium3}). As expected, $\left\langle
E_{B}\right\rangle $ is the same as the binding energy determined in Ref.
\onlinecite{Haziot13c}, where the distribution of $E_{B}$ was not considered.
Our best fit value of $\left\langle E_{B}\right\rangle $ also shows that our
determination of $F_{c}$ (Fig. \ref{fig:temperature-dependence}) is
reasonable, since $F_{c}$ is the maximum magnitude of the spatial derivative
of $E_{B}$ where the $^{3}$He atom is bound and it has a numerical value
$\approx\left\langle E_{B}\right\rangle /4b$. The quality of the fits in Fig.
\ref{fig:helium3} is better than that of the fits to $\mu\left(  T\right)  $
and $Q^{-1}\left(  T\right)  $ in Ref. \onlinecite{Kang13b}. In that work,
unlike the present work, the frequency dependence was not studied and the fits
in the $^{3}$He binding regime were not constrained by fits to measurements of
phonon damping.

\section{Discussion and Conclusion}

As discussed in Ref. \onlinecite{Haziot13c}, there is a critical dislocation
speed below which $^{3}$He atoms move with the dislocations and damp their
motion. Above the critical speed, $^{3}$He atoms approximate static pinning
sites. We established above that the measurements in Fig. \ref{fig:helium3}
were made below the critical dislocation speed. Some previous frequency
dependent measurements of $\mu\left(  T\right)  $ and $Q^{-1}\left(  T\right)
$ in $^{4}$He polycrystals were interpreted in terms of a Debye model with a
distribution of activation energies.\cite{Syshchenko10} The equations that
yielded the best fits to $\mu\left(  T\right)  $ and $Q^{-1}\left(  T\right)
$ in that work are equivalent to the ones used to calculate $\mu\left(
T\right)  $ and $Q^{-1}\left(  T\right)  $ in Fig. \ref{fig:helium3} in the
limit of small softening $\Lambda L_{N}^{2}$ and a single dislocation network
length. The standard deviation of the distribution of binding energies that
was determined in Ref. \onlinecite{Syshchenko10} is 2.5 times larger than in
the present work, perhaps because the distribution of network lengths was not
considered in Ref. \onlinecite{Syshchenko10}.

The resonant period of a torsional oscillator containing solid helium
decreases with the shear modulus of the helium.\cite{Maris11} Many if not all
of the torsional oscillator results initially explained by supersolidity of
$^{4}$He can be explained by this effect.\cite{Maris12,Beamish12} In Ref.
\onlinecite{Iwasa13}, results of torsional oscillator experiments were
explained in terms of $^{3}$He pinning of dislocations. This explanation could
be consistent with the model used in the present work if the dislocations were
moving faster than the critical dislocation speed in the torsional oscillator
measurements that were analyzed. To verify this, it would be necessary to
study the frequency dependence of a torsional oscillator measurement at
constant response amplitude, analogously to Fig. \ref{fig:Arrhenius}. In Ref.
\onlinecite{Iwasa13}, the distribution of network lengths was determined, up
to the critical force, from the dependence of the period shift on the rim
speed in the torsional oscillator experiment of Ref. \onlinecite{Aoki07}. Thus
the method for determining the form of the network length distribution was
analogous to the one used in the present work. However, it was apparently
impossible to observe phonon damping in the torsional oscillator measurements analyzed in Ref. \onlinecite{Iwasa13}, and the
critical force was left as a free parameter in the fit to the temperature
dependence of the period shift, unlike the fit to the temperature dependence
of $\mu$ and $Q^{-1}$ in Fig. \ref{fig:helium3} of the present work. After
optimizing the value of the critical force, the best fit period shift as a
function of temperature was a factor of 1.45 below the measured period shift
in Ref. \onlinecite{Iwasa13}.

In conclusion, we used the unique properties of helium crystals to measure the
length distribution of a dislocation network. We showed that it was necessary
to account for this broad distribution to obtain a complete, consistent and
quantitative interpretation of the mechanical properties of these crystals as
a function of temperature, amplitude and frequency of driving strain in both
the phonon and $^{3}$He damping regimes. In so doing, we obtained detailed
information about the interactions between the dislocations and $^{3}$He
impurities. We hope that this work inspires calculations of the $^{3}$He
binding energy as a function of the screw/edge character of the dislocation to
which the $^{3}$He atom is bound.

\section{ACKNOWLEDGMENTS}

This work was supported by grant ERC-AdG 247258 SUPERSOLID, and by a grant
from NSERC Canada.

\section{Appendix A: Crystal Orientation Dependence}

For clarity of presentation in the equations of the main text, we neglected
the small angle between our crystal's six fold axis of symmetry and the $z$
direction defined in Fig. \ref{fig:mu(epsilon)}, but we accounted for this
small angle in our calculations. The elastic coefficients of a hcp crystal can
be labelled using Voigt notation, where the subscripts $1,2,3,4,5,6$
correspond to the coordinates $x^{\prime}x^{\prime},y^{\prime}y^{\prime
},z^{\prime}z^{\prime},y^{\prime}z^{\prime},x^{\prime}z^{\prime},x^{\prime
}y^{\prime}$ in a Cartesian coordinate system where $z^{\prime}$ is aligned
with the six-fold axis of symmetry (the elastic coefficients are invariant
under rotations about $z^{\prime}$). In hcp $^{4}$He, $c_{44}$ is the only
elastic coefficient that is reduced by dislocation glide,\cite{Haziot13a} so
that $\epsilon_{4}=\left(  \epsilon_{\text{dis}}+\epsilon_{\text{el}}\right)
$. Since $c_{44}=\sigma_{4}/\epsilon_{4}$, we can replace Eq. \ref{eq:mudef}
by%
\begin{equation}
c_{44}=\frac{c_{44}^{\text{el}}}{1+\epsilon_{\text{dis}}/\epsilon_{\text{el}}%
}\label{eq:c44def}%
\end{equation}
where $\epsilon_{\text{el}}=\sigma_{4}/c_{44}^{\text{el}}$ and $c_{44}%
^{\text{el}}=124$ bar is the value of $c_{44}$ in the absence of mobile
dislocations at our working pressure of 25.3 bar.\cite{Greywall77} The stress
$\sigma$ must be replaced by $\sigma_{4}$ in Eqs. \ref{eq:ofmot} and
\ref{eq:xi_sig}, but we note that Eq. \ref{eq:edis_gen} remains the same due
to cancellation of the factor $\sigma_{4}$. In order to obtain the theoretical
curves in the figures of the main text, we substituted Eq. \ref{eq:c44def}
into
\begin{equation}
\mu=0.97c_{44}+0.03c_{66}^{\text{el}},\label{eq:muofc44}%
\end{equation}
where $c_{66}^{\text{el}}=96.0$ bar. Equation \ref{eq:muofc44} follows from
the orientation our crystal (Sec. \ref{sec:Experiment}) and the general
expression for $\mu$ given in the supplement to Ref. \onlinecite{Haziot13a}
for arbitrary crystal orientation.

To fit the measurement of $\mu(\epsilon)$ shown in blue in
Fig.~\ref{fig:distribution}, we substituted the fitting function:
\begin{equation}
\frac{c_{44}}{c_{44}^{\text{el}}}=\frac{\tanh{[c_{1}(\ln{L_{c}}+c_{0})]}%
}{c_{2}}+c_{3}%
\end{equation}
into Eq. \ref{eq:muofc44} and obtained best fit values $c_{0}$ = 9.05, $c_{1}$
= 1.83, $c_{2}$ = 3.36, and $c_{3}$ = 0.703.

\section{Appendix B: Arrhenius Equation}

Substituting Eq. \ref{eq:edis_sing_3He} into Eq. \ref{eq:mudef} yields%
\begin{equation}
\frac{\mu}{\mu_{\text{el}}}=\frac{1+s+\left(  \omega\tau_{3}\right)
^{2}+i\omega\tau_{3}s}{\left(  1+s\right)  ^{2}+\left(  \omega\tau_{3}\right)
^{2}}%
\end{equation}
where $s\equiv\alpha\Lambda L_{N_{0}}^{2}$ and $\tau_{3}=B_{3}L_{N}^{2}%
/\pi^{2}C$. Since $Q^{-1}=\operatorname{Im}[\mu]/\operatorname{Re}[\mu]$, we
have
\begin{equation}
Q^{-1}=\frac{\omega\tau_{3}s}{1+s+\left(  \omega\tau_{3}\right)  ^{2}}.
\end{equation}
The dissipation attains a maximum value $Q_{\max}^{-1}=s/2\sqrt{1+s}$ for
$\omega\tau_{3}=\sqrt{1+s}$. We defined the temperature at which $Q^{-1}$ is
maximized as $T_{p}$ in the main text. Thus%
\begin{equation}
\omega\tau_{0}\exp\left[  E/T_{p}\right]  =\sqrt{1+s} \label{dissmax}%
\end{equation}
where $\tau_{0}=B_{3}^{\infty}L_{N}^{2}/\pi^{2}C$. Rearranging Eq.
\ref{dissmax} yields Eq. \ref{eq:Arrhen_law}.

\bibliography{plasticity-references}

\end{document}